\documentclass[11pt]{revtex4-1}
\usepackage[latin9]{inputenc}
\usepackage{geometry}
\usepackage{array}
\usepackage{float}
\usepackage{textcomp}
\usepackage{amsbsy}
\usepackage{amssymb}
\usepackage{graphicx}
\usepackage{esint}
\usepackage{units}

\makeatletter

\@ifundefined{textcolor}{}
{%
 \definecolor{BLACK}{gray}{0}
 \definecolor{WHITE}{gray}{1}
 \definecolor{RED}{rgb}{1,0,0}
 \definecolor{GREEN}{rgb}{0,1,0}
 \definecolor{BLUE}{rgb}{0,0,1}
 \definecolor{CYAN}{cmyk}{1,0,0,0}
 \definecolor{MAGENTA}{cmyk}{0,1,0,0}
 \definecolor{YELLOW}{cmyk}{0,0,1,0}
}

\usepackage[font=small]{caption}
\usepackage{epstopdf}
\usepackage{dcolumn}
\usepackage{bm}
\usepackage{array}
\usepackage{tabto}

\usepackage{color}

\makeatother

\begin{document}

\title{Optical probing of the intrinsic spin Hall effect in a high mobility GaAs p-doped quantum well}

\author{A. Maier$^{1}$,}

\email{adrmaier@phys.ethz.ch}

\author{C. Reichl$^{1}$, S. Riedi$^{1}$, S. Faelt$^{1}$ and W.
Wegscheider$^{1}$}

\affiliation{$^{1}$Solid State Physics Laboratory, ETH Zurich, Otto-Stern-Weg
1, 8093 Z\"urich, Switzerland}

\begin{abstract}

We report on the detection of the intrinsic spin Hall effect in a modulation doped AlGaAs/GaAs/AlGaAs heterostructure bounded by a self-aligned pn-junction, fabricated by the cleaved edge overgrowth method. Light emission due to the recombination of electrons and spin-polarized holes was generated and mapped with a spatial resolution of one micrometer. An edge accumulated spin polarization of up to $\unit[11]{\%}$ was measured, induced solely by application of an electric field perpendicular to the pn-junction. Using a quantum dot structure as light source, a linear dependence of the effective spin polarization, and with that the dominance of the spin Hall effect, with the electric field is seen. Spatially resolved spectroscopy from an epitaxially fabricated LED is demonstrated to be a valuable tool to probe the edge states of electron and hole gases in reduced dimensions. 

\end{abstract}
\maketitle

\section{Introduction}
\label{sec:intro}

The field of spintronics relies on the ability to control the electron's spin degree of freedom in semiconductor materials \cite{Wolf2001}. Spin-orbit interaction (SOI) provides the basis for such a control mechanism. The spin Hall effect (SHE) results in a spin current perpendicular to an applied electrical field leading to spin separation and accumulation at the edges of a sample. In contrast to the standard Hall effect, the SHE occurs without applying an external magnetic field. 

In the emergence of the SHE, two conceptually different mechanisms can be distinguished. First, the extrinsic SHE arises from spin-dependent scattering off static impurities present in the sample investigated \cite{Dyakonov1971, Hirsch1999, Zhang2000}. Second, there is the so called intrinsic SHE, where the deflection of spin carriers stems from SOI with the crystal lattice \cite{Murakami2003, Sinova2004, Schliemann2005}. Here the two contributors are bulk inversion asymmetry in the form of Dresselhaus SOI \cite{dresselhaus1955} and structural inversion asymmetry termed Rashba SOI \cite{Bychkov1984}. The latter can be set by designing the semiconductor structure appropriately as to result in an asymmetric electrical field, or by applying a gate voltage to tune the electrical field seen by the spin carriers and thereby the strength of the Rashba SOI.

To investigate the SHE in two dimensional electron and hole systems (2DES and 2DHS), the technique of molecular beam epitaxy (MBE) in the gallium arsenide/aluminium gallium arsenide (GaAs/AlGaAs) material system lends itself due to the very precise growth process and the exceptionally high quality of the structures \cite{Reichl2014, Manfra2014, Hirmer2013, Watson2012}. 

The electrically induced spin accumulation in GaAs was first observed in an n-doped bulk system by means of Kerr-rotation measurements and was found to be of the extrinsic type \cite{Kato2004}. Shortly thereafter the same group reported the observation of the extrinsic SHE in a 2DES \cite{Sih2005}. At the same time Wunderlich \textit{et al.} discovered the intrinsic SHE in a 2DHS with triangular confinement potential \cite{Wunderlich2005}. In this work a sophisticated lithographic technique had to be employed to match the region of spin accumulation with that where charge carrier recombination occurs. However, as a result of the small extension of the spin accumulation region, the effectiveness of this approach can be debated. While it is still an ongoing topic with a recent publication showing intrinsic SHE in a multilayer 2DES \cite{Hernandez2013}, there are only few findings of the intrinsic SHE especially in 2DHSs.

Here we report on the detection of the intrinsic SHE in a single sidedly p-doped rectangular quantum well. Spin accumulation was generated by applying an electric field as a result of which the SHE creates an imbalance of out of plane hole spins at opposite edges of the sample. In contrast to previous work, a lateral pn-junction, which is self-aligned to the 2DHG boundary, was created to optically probe the edge spin signal, using the technique of cleaved edge overgrowth (CEO). A comparative analysis of magnetotransport characteristics suggests that the SHE is of the intrinsic type. Furthermore an impurity based quantum dot structure with a size below the diffraction limit of our detection system shows high spin polarization which is dependent on the electric field applied.

\section{Experimental Details}
\label{sec:experiment}

\begin{figure}
\centering
\includegraphics[width=\textwidth]{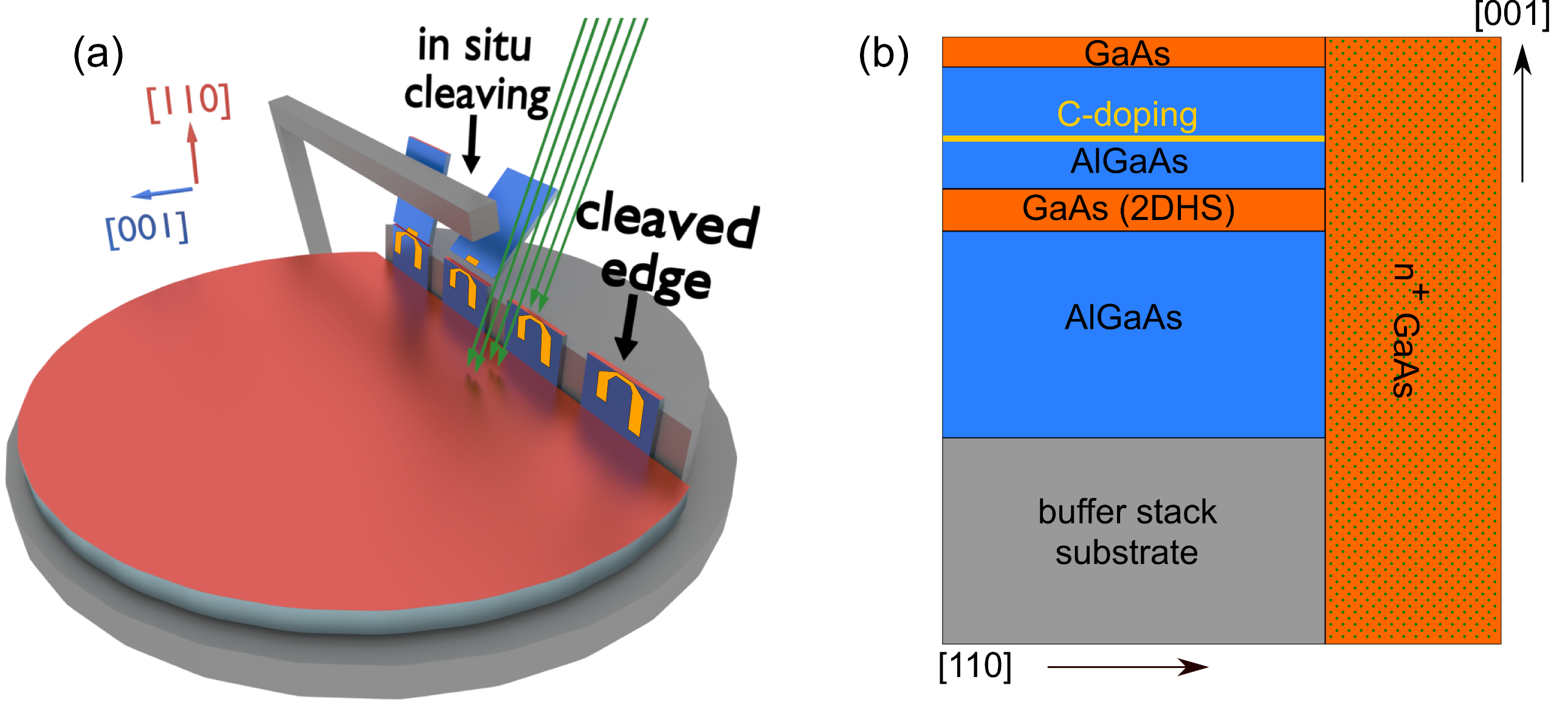}
\vspace*{-0.3cm}
\caption{
(a) The CEO technique: A rotating metal bar is used to cleave preprocessed samples in the UHV of the growth chamber. The (110) surface is then immediately overgrown with a second heterostructure (green arrows). (b) Cut through the final structure. At the interface of 2DHS and n$^+$ GaAs the lateral pn-junction is formed.
}
\label{fig:CEO}
\end{figure}

In order to gain information on the degree of spin accumulation at the edges of the sample a lateral pn-junction \cite{Miller1985} was fabricated utilizing the CEO technique \cite{Pfeiffer1990, Pfeiffer1993} (Fig. \ref{fig:CEO}a). In contrast to previous attempts along this line \cite{Wunderlich2005} in this way a minimally disturbing atomically sharp edge is self-aligned with a light emitting diode.
The basis for the sample is comprised of a $\unit[15]{nm}$ wide Al$_{0.3}$Ga$_{0.7}$As/GaAs/Al$_{0.3}$Ga$_{0.7}$As quantum well, grown on a (001) GaAs substrate by the method of MBE. Population of the 2DHS was achieved by single side p-doping. The resulting electric field tilts the confinement potential of the quantum well and therefore enhances Rashba spin-orbit interaction \cite{giglberger2007}. Magnetotransport characterizations performed on a $\unit[4 \times 4] {mm^2}$ sample show a high carrier mobility of $\mu=\unit[2.3 \times 10^5]{cm^2}V^{-1}s^{-1}$ and a hole density of $p=\unit[3.2 \times 10^{11}]{cm^{-2}}$ at a temperature of $\unit[4.2]{K}$.

This structure was processed by optical photolithography and chemical wet etching into a $\unit[100]{\mu m}$ long and $\unit[20]{\mu m}$ wide mesa with leads. Then, a scratch on the edge of $\unit[7 \times 7]{mm^2}$ square shaped pieces of the wafer was applied using a tungsten carbide needle. This scratch defined a predetermined breaking line and was aligned to cut the mesa in half along its length axis. After cleaning, the samples were mounted on a special substrate holder  and introduced into the ultra high vacuum (UHV) of the MBE growth chamber. A rotating metal bar was used to cleave the samples along the predefined line given by the scratch. This way, an atomically flat (110) surface was exposed and immediately overgrown with $\unit[500]{nm}$ n$^+$ bulk GaAs, resulting in the structure depicted in Fig. \ref{fig:CEO}b. Details of the procedure preceding the cleaving process can be found in \cite{Riedi2016}.

The complete sample consists of a $\unit[10]{\mu m}$ wide channel which is self aligned to a lateral pn-junction at the interface of the n-doped GaAs and the 2DHS. Note, however, that the effective channel width for an applied current is smaller than the mesa. The difference in dimensions of the overgrown bulk n-GaAs and the 2DHS leads to an asymmetric space charge region depleting a substantial part of the channel.
Ohmic contacts to the 2DHS were applied by soldering and annealing of an alloy of $\unit[96]{\%}$ Indium and $\unit[4]{\%}$ Zinc (by weight).

\begin{figure}
\centering
\includegraphics[width=\textwidth]{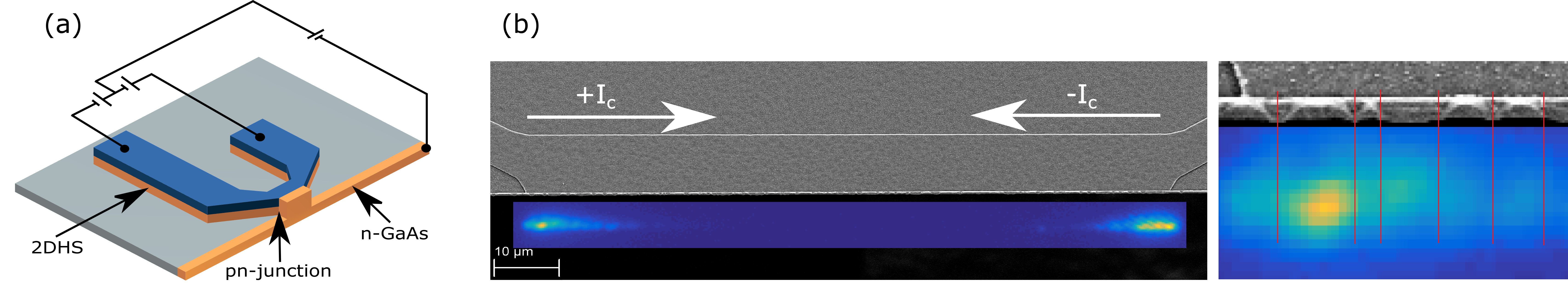}
\vspace*{-0.3cm}
\caption{
(a) Sample design and three terminal electrical setup for changing longitudinal current and bias of the pn-junction independently. (b)SEM image of the final lateral pn-junction with directions of longitudinal current $I_c$ for orientation. The $\unit[500]{nm}$ n$^+$ bulk GaAs region appears at the lower edge of the sample as a thin bright line. Colored insert shows typical emission intensity. Note that it is a stitched image of two measurements with opposing current directions. The decrease in intensity due to the voltage drop along the channel can cleary be seen. The magnified cutout shows a closeup of intensity oscillations due to varying signal dampening of the CEO facets. 
}
\label{fig:channel}
\end{figure}

An electric three terminal setup was chosen to be able to drive a current through the channel and vary the bias of the LED formed by the lateral pn-junction (Fig. \ref{fig:channel}a) independently. The current through the mesa can be driven in both directions (designated as positive and negative $I_c$) to accumulate either of both spin flavours at the pn-junction. 
All measurements regarding the analysis of the edge spin accumulation were carried out using a liquid helium flow cryostat with a window for optical access and piezo actuators to enable spatial mapping.

A confocal optical setup provided a diffraction limited spatial resolution of $\sim \unit[1]{\mu m}$.
Spectral analysis of both photoluminescence and electroluminescence emanating from the sample was carried out by the use of a CCD spectrometer.
A He-Ne laser with a wavelength of $\unit[633]{nm}$ provided the excitation energy for photoluminescence (PL) measurements.
Analysis of the circular polarization of the emitted light was performed by using a combination of a half-wave plate, a quarter wave-plate, a polarizing beam splitter and a CCD sensor for synchronous detection.
Polarization was then determined by calculating the weighted difference of intensities $I_1$ and $I_2$ of both polarized beams: $P = (I_1-I_2) / (I_1+I_2)$.

\section{Results and Discussion}
\label{sec:results}

After cooldown the contacts to the 2DHS show ohmic behaviour. The lateral pn-junction exhibits clearly the rectifying characteristic of a diode, emitting light when biased in forward direction. However, despite the high mobility of the 2DHS the channel shows a significant resistivity. Two point measurements yielded a value of $R_c \sim \unit[30]{k\Omega}$ at low temperatures. The resulting potential drop under current flow along the channel leads to a decrease in PL intensity away from the corners of the LED. Thus only the first few $\mu m$ of the mesa emit light and the regions can be treated as individual LEDs with no overlap.

\begin{figure}
\centering
\includegraphics[width=\textwidth]{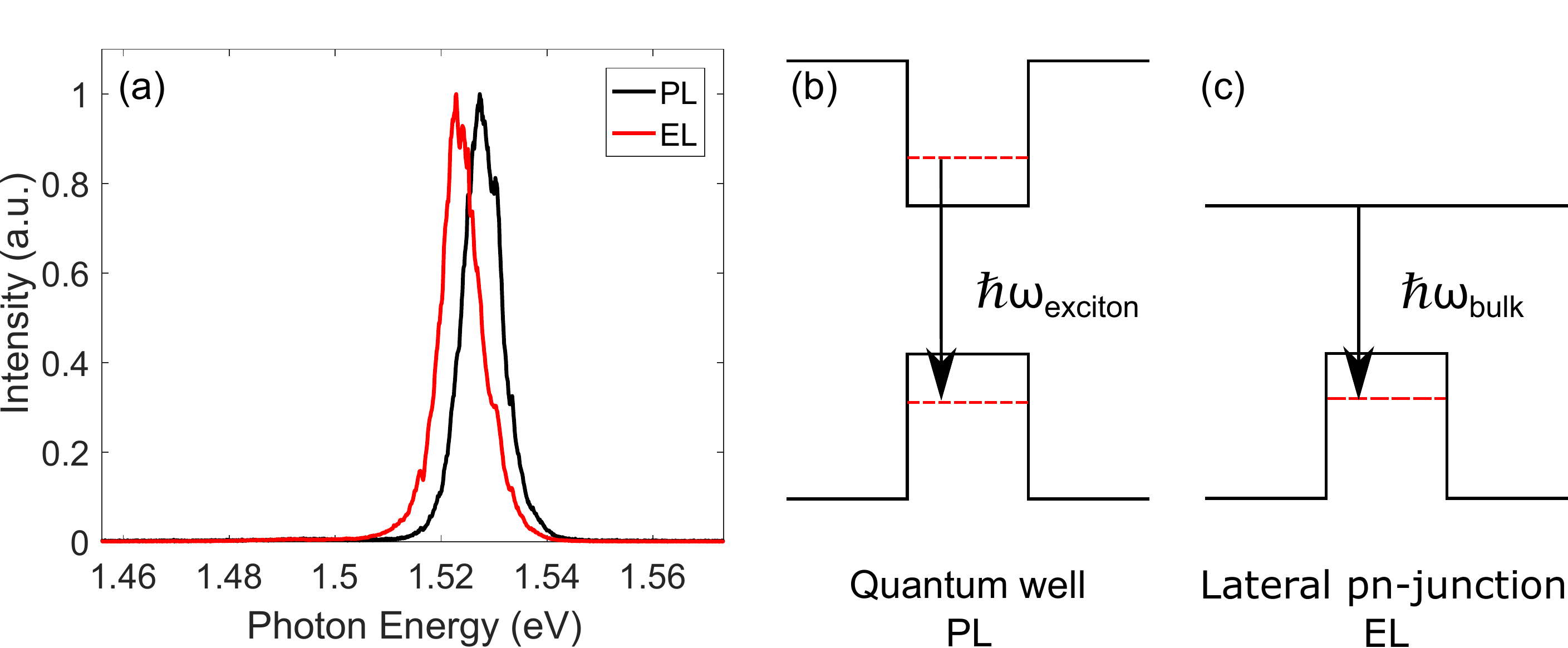}
\vspace*{-0.3cm}
\caption{
(a) Spectrum of PL in the middle of the mesa and EL at the lateral pn-junction. (b) Photon energy of an exciton is given by the confinement potential and the exciton binding energy. (c) At the lateral pn-junction recombination of bulk electrons and 2D holes takes place. Since there is no confinement potential in the conduction band the energy of EL photons is lower compared to PL photons.
}
\label{fig:spectrum}
\end{figure}

The spectral peak of PL measurements of the quantum well region at an energy of $\unit[1.527]{eV}$ (Fig. \ref{fig:spectrum}a) can thus be related to the heavy-hole exciton \cite{Harrison2016}. The electroluminescence (EL) peak from the CEO interface of 2DHS and bulk GaAs on the other hand was found to be at $\unit[1.522]{eV}$. This value corresponds to theoretically expected transition energies obtained from the 8-band Schr\"odinger-Poisson-solver software Nextnano \cite{Nextnano}: The simulation of the actual sample structure shows that the lack of a confinement potential in the conduction band leads to a redshift of EL at the lateral pn-interface compared to the PL signal by $\unit[5]{meV}$.

The tight binding model describes heavy holes (HH) and light holes (LH) by p-like orbitals with a total angular momentum $j=3/2$. LH are made of $m= \pm 1/2$ states whereas HH are made of $m= \pm 3/2$ with the corresponding quantization axis parallel to the quantum well growth direction [001]. Thus HH spins are aligned perpendicular to the 2D system they reside in \cite{winkler2008}. Furthermore in a 2DHS the degeneracy of HH and LH at $k=0$ is lifted and HH-LH splitting occurs. The observed and the theoretical values confirm that for the given hole density only the lowest HH subband is contributing to the 2DHS. Both the LH and the split off subband are not populated.

For the recombination of HH and electrons optical selection rules dictate either a $\sigma^+$ or $\sigma^-$ circular polarization of emitted light depending on the spin of the HH involved.
As a consequence a predominant spin direction, which accumulates at the interface of the 2DHS and the n$^+$ GaAs because of the SHE, can be probed by analyzing the degree of circular polarization emitted from there.

\begin{figure}
\centering
\includegraphics[width=\textwidth]{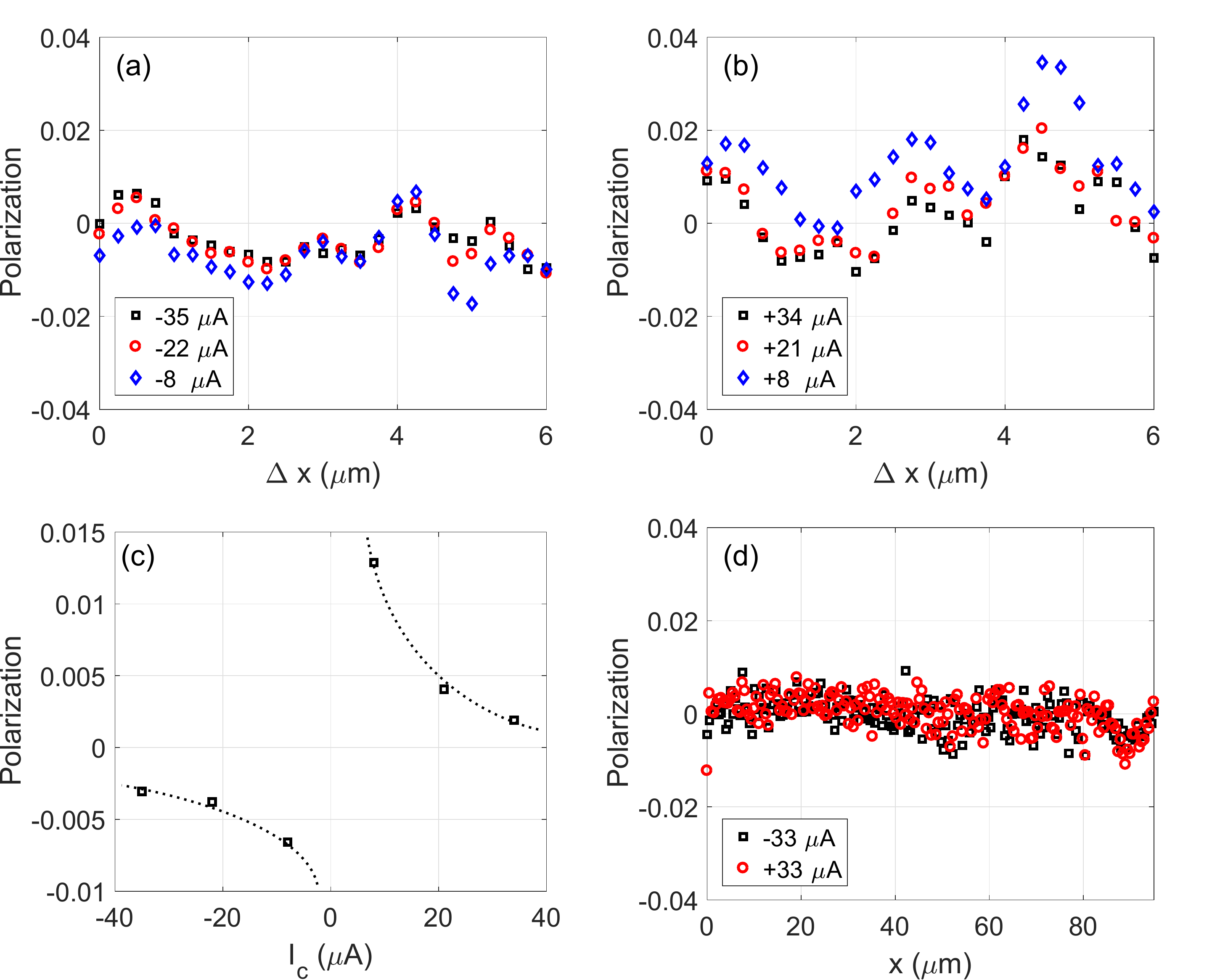}
\vspace*{-0.3cm}
\caption{
Degree of circular polarization of light emitted by the lateral pn-junction. Polarization for different negative (a) and positive (b) currents driven through the channel. Note that $\Delta x$ is defined as distance from start of emission in each direction of current. The higher voltage for higher currents leads to larger recombination zones reaching away from the edge into the bulk of the channel where no spin accumulation can occur. The increased ratio of unpolarized hole spins explains the counterintuitively lowered polarization for higher currents. (c) Mean values of polarization as a function of current through the 2DHS channel. Dotted lines serve as guide to the eye. (d) For reference a 2DES p$^+$ interface was also examined. The lateral np-junction shows emission along the whole length of the channel. No change in circular polarization can be observed for opposing currents. Here, $x$ denotes the same absolute position for both measurements.
}
\label{fig:polarization}
\end{figure}

In Fig. \ref{fig:polarization}a we show the progression of the degree of circular polarization measured along the lateral pn-junction for negative currents $I_c$. The quantity $\Delta x$ is defined as the distance from the starting point of the interface in the direction of $I_c$. For comparison three different currents were applied along the channel while the bias across the LED was altered to ensure a constant $I_{pn} = \unit[2.5]{\mu A}$. While circular polarization shows some variation depending on distance, a shift to lower polarization can be noticed when lowering the absolute current $|I_c|$.

A more pronounced picture presents itself for positive $I_c$ (Fig. \ref{fig:polarization}b). While there is also a distance dependent variation in polarization, it now increases significantly for lower $|I_c|$. This behaviour contradicts the expected linear relation of spin polarization and longitudinal electrical field in the channel $S \propto E_c$ \cite{nomura2005b}. We attribute this to the small spin accumulation region at the edge where its size is defined by the spin-orbit length $L_{SO}$.

Magnetotransport measurements of structures produced with layer stacks and growth parameters similar to the sample presented here yield $L_{SO} \sim \unit[100]{nm}$ \cite{Nichele2014}. However, we found the recombination zone sizes of HH and electrons to be larger than $\unit[1]{\mu m}$ for higher absolute currents. This implies a high ratio of mixing of unpolarized bulk HH to polarized edge accumulated HH reducing the observable degree of circular polarization. 

Only for $I_c=\pm \unit[8]{\mu A}$ the recombination zone was not distinguishable from the diffraction limit of the optical setup, thus increasing the polarization detected. However, still a significant mixing with unpolarized HH is expected and only probing the pure spin accumulation region would yield the true spin polarization. Note that the same issue arose in the first experiment to detect the intrinsic SHE \cite{Wunderlich2005}.
Direct comparison of $I_c= \pm \unit[8]{\mu A}$ shows a SHE induced mean difference of $\unit[2]{\%}$ in circular polarization and a change in sign for opposing current directions with polarization values of $\unit[-0.7]{\%}$ and $\unit[+1.3]{\%}$ respectively (Fig. \ref{fig:polarization}c).
The apparently oscillatory behaviour of the polarization is an artifact caused by irregular faceting which occurs when the (001) crystal plane meets with the (110) overgrowth layer: While overgrowth of the cleaved surface at the 2DHS matches exactly the underlying crystal structure, at edges this process leads to formation of facets of variable sizes which partly cancel the polarization before the signal reaches the detector. The close up in Fig. \ref{fig:channel}b shows these facets and their alignment with the signal strength.

Calculations derived from the aforementioned study of transport properties of similar samples help to determine the type of SHE observed. The ratio of the energy associated with momentum scattering and Fermi energy $(\hbar / \tau)/ E_F \approx 0.01 $ and spin-orbit splitting energy to Fermi energy $\Delta_{SO}/E_F \approx 0.5$ suggest the sample to be in the weak disorder, strong spin-orbit coupling regime. With the robustness of the effect shown \cite{Hankiewicz2004, nikolic2005, nomura2005d} we attribute our results to stem from the intrinsic SHE.

For comparison, we have performed the same experiment on a 2DES structure that was overgrown with p$^+$-doped GaAs, thus forming an np-junction. The 2DES resides in a single sidedly n-doped, $\unit[15]{nm}$ wide quantum well. It has a similar density of $n=\unit[2 \times 10^{11}]{cm^{-2}}$, and a mobility of $\mu=\unit[2 \times 10^6]{cm^2}V^{-1}s^{-1}$, which is a rather low value for GaAs-based electron systems. However, it surpasses our 2DHS' mobility by an order of magnitude, leading to lower resistivity and with that recombination along the whole length of the np-junction. Spin accumulation was probed with a current of $\pm \unit[33]{\mu A}$. With a resistivity of $R_c \sim \unit[10]{k\Omega}$, this measurement corresponds to the measurement with $I_c = \pm \unit[8]{\mu A}$ conducted on the 2DHS.
As expected, spin polarization effects of the electron system cannot be observed or lie within the margin of error for this measurement (Fig. \ref{fig:polarization}d). This compares to other measurements of 2DES where only a small degree of polarization could be observed \cite{Sih2005}, and emphasizes the advantage of employing a 2DHS for the utilization of SOI for spintronics.

\section{Dot-Like Emitter}

\begin{figure}
\centering
\includegraphics[width=\textwidth]{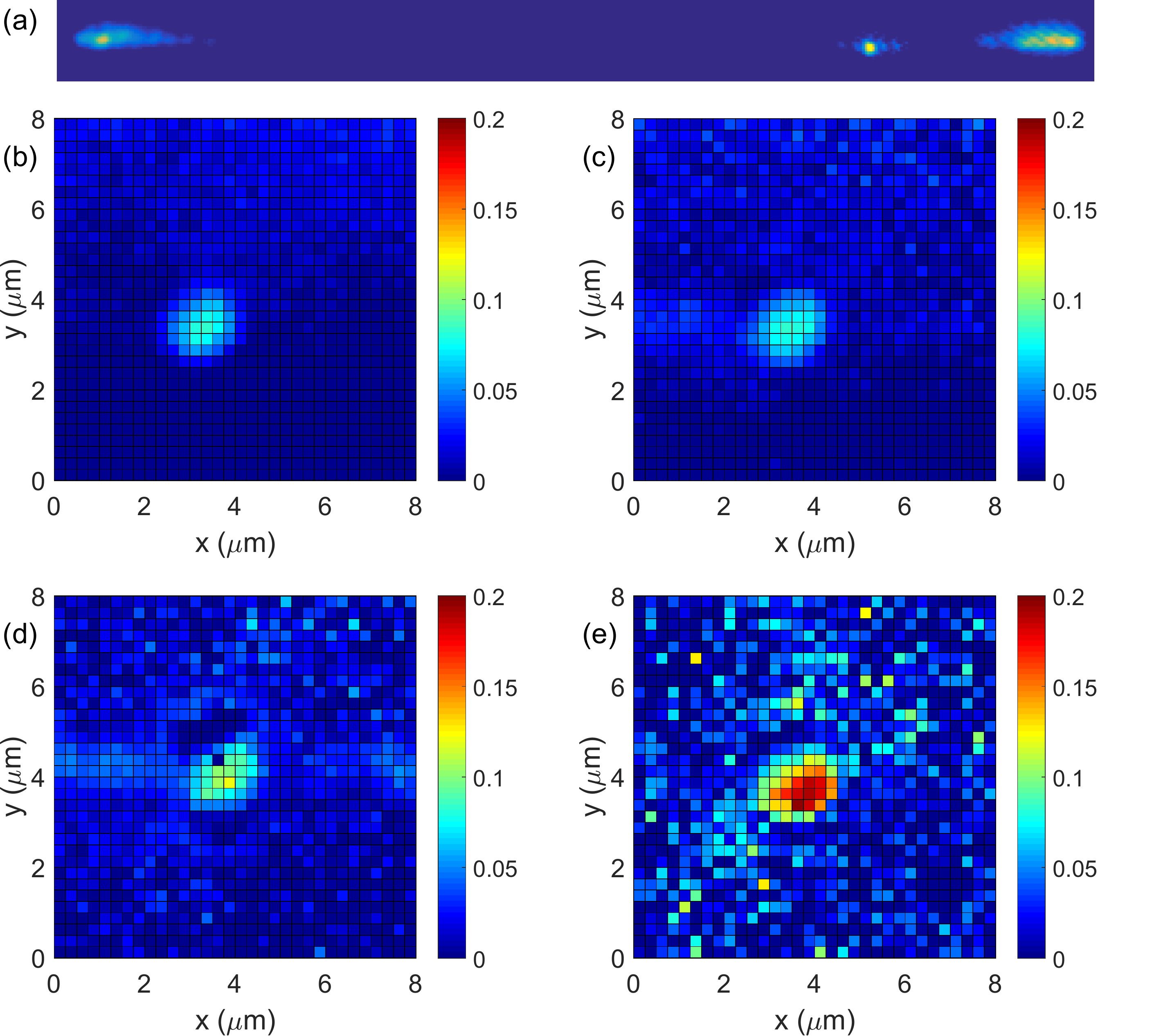}
\vspace*{-0.3cm}
\caption{
(a): Position of the dot emitter on the mesa. (b) to (e): Evolution of polarization at the recombination center for different currents along the channel: $I_c=\unit[-37]{\mu A}$ (b), $I_c=\unit[-22]{\mu A}$ (c), $I_c=\unit[5]{\mu A}$ (d) and $I_c=\unit[22]{\mu A}$ (e). Note the interference pattern visible in (c) and (d) indicating a size of the active recombination region below the diffraction limited resolution of the optical setup. 
}
\label{fig:polspot}
\end{figure}

As described above, the pn-junction exhibits emission only near the corners due to the potential drop along the 2DHS channel. However, at a position of $\unit[80]{\mu m}$ emission can be observed for a wide range of applied longitudinal currents (see Fig. \ref{fig:polspot}a). We performed a spatial mapping of the polarization of the recombination center for different $I_c$ (Fig. \ref{fig:polspot}b - e). The constant current over the lateral pn-junction was chosen to be $I_{pn} = \unit[10]{\mu A}$.  The size of the recombination region appears to be $\sim \unit[1]{\mu m}$. However, since an interference pattern (best seen in Fig. \ref{fig:polspot}d) emerges we conclude that the actual size is far below the diffraction limit and we refer to the region as dot-like.

\begin{figure}
\centering
\includegraphics[width=\textwidth]{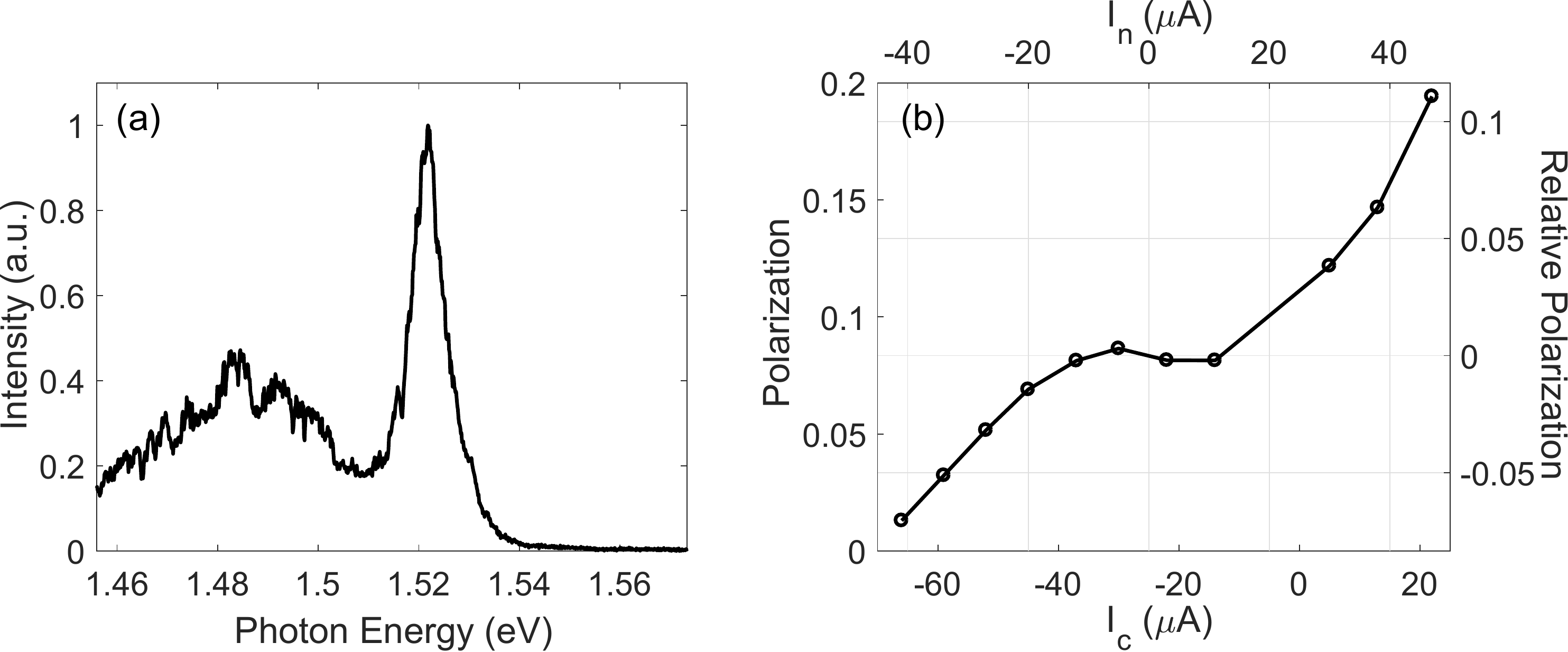}
\vspace*{-0.3cm}
\caption{
(a) Spectrum of the recombination center showing the EL peak at an energy of $\unit[1.517]{eV}$ and a broad signal from an impurity at a lower energy of $\unit[1.49]{eV}$, which is attributed to a residual carbon atom. Integration of the spectrum yields a four times higher total intensity of the impurity region compared to the EL peak. (b) Polarization of the dot-like emitter as a function of current through the 2DHS channel. Top and right axis showing offset-adjusted current and polarization due to the position and the inherent anisotropy of the dot.
}
\label{fig:polcurrent}
\end{figure}

Spectral analysis of the dot-like emitter shows an additional broad emission around an energy of $\unit[1.49]{eV}$ (Fig. \ref{fig:polcurrent}a). Despite the very high purity of the structures produced in our MBE setup, a low background of residual atoms is always incorporated during growth. Earlier investigations performed in our group showed that this impurity background consists predominantly of carbon compounds, and a luminescence peak associated to carbon and matching our measurements can be found in a variety of samples \cite{Reichl2014, Schlaepfer2016}. 

The small dimensions of the dot-like emitter and its position at the CEO-interface turns it into an appropiate tool to probe the degree of spin accumulation at the edge of the 2DHS for an extended range of longitudinal currents where emission is observable at the same position. A total variation in polarization of $\unit[18]{\%}$ can be generated for currents rating from $\unit[-66]{\mu A}$ to $\unit[+22]{\mu A}$ (Fig. \ref{fig:polcurrent}b).

The measured polarization shows linear behaviour for positive and large negative currents, fitting $S \propto E_c$, although a saddlepoint around $I_c=\unit[-25]{\mu A}$ was found. This feature is due to the fact that for $I_{pn} > |I_c|$ current is injected in both sides of the channel to fulfill the condition of a constant current across the pn-junction. Thus, a strongly inhomogeneous $E_c$ is expected in the progression of the channel, greatly disturbing spin accumulation by the SHE. Taking also into account the off-center position of the dot-like emitter, a current offset for the unpolarized state can be anticipated.
Lastly the dot-like emitter shows finite polarization for all currents and therefore exhibits no change in sign of the polarization. We relate this characteristic to an anisotropic nature of the impurity at the interface. Polarization anisotropies have been studied for various systems like quantum dots, quantum wells and especially vertical-cavity surface-emitting lasers where they are used to facilitate a specific polarization of light emission \cite{gammon1996, vankesteren1990a, blackwood1994, vanexter1998}.
As a result of these considerations we introduce a net current $I_n = I_c + \unit[25]{\mu A}$ and a relative polarization with an offset of $\unit[-8.3]{\%}$ (Fig. \ref{fig:polcurrent}b) top and right axis). We conclude the dot-like emitter to show the SHE induced spin accumulation for net currents $I_n < \unit[-10]{\mu A}$ with a negative polarization sign and for currents $I_n > \unit[10]{\mu A}$ with a positive polarization sign. The maximally achievable values were $\unit[-7]{\%}$ and $\unit[11]{\%}$ respectively. Being an order of magnitude higher compared to the measurements presented before, this indicates that recombination does not include unpolarized bulk HH and thus only the small accumulation region within $L_{SO}$ is probed.

\section{conclusion}

The SHE was observed in a single sidedly p-doped Al$_{0.3}$Ga$_{0.7}$As/GaAs/Al$_{0.3}$Ga$_{0.7}$As quantum well. Employing the CEO-technique, a lateral pn-junction was attached to the (110) edge of the 2DHS to probe spin accumulation of HH. Analysis of the circular polarization of light emitted at the lateral LED shows a change in sign and values of $\unit[-0.7]{\%}$ and $\unit[+1.3]{\%}$ of the measured polarization for both current directions. This is in agreement with the results of Wunderlich et al. \cite{Wunderlich2005}, where a triangular 2DHS showed a similar degree of polarization. A comparison to similarly grown structures indicates strong SOI and weak disorder and the SHE to be of the intrinsic type.
An impurity enhanced dot-like emitter at the lateral pn-junction was investigated. Its size, presumably smaller than the spin accumulation region, allowed us to measure the pure spin accumulation region showing an achievable polarization of up to $\unit[11]{\%}$, almost an order of magnitude higher than measured at other positions of the lateral pn-junction. To our knowledge, this is the highest value reported so far for an electrically induced intrinsic SHE. In general the CEO fabricated LED proved to be an appropiate tool to probe edge states of 2D hole and electron systems.

\section{Acknowledgments}
The authors thank Lars Tiemann and Martin Kroner for stimulating discussions.
This work was supported by the Swiss National Science Foundation (Schweizer Nationalfonds, NCCR "Quantum Science and Technology").

\bibliography{Paper}

\begin{thebibliography}{35}%
\makeatletter
\providecommand \@ifxundefined [1]{%
 \@ifx{#1\undefined}
}%
\providecommand \@ifnum [1]{%
 \ifnum #1\expandafter \@firstoftwo
 \else \expandafter \@secondoftwo
 \fi
}%
\providecommand \@ifx [1]{%
 \ifx #1\expandafter \@firstoftwo
 \else \expandafter \@secondoftwo
 \fi
}%
\providecommand \natexlab [1]{#1}%
\providecommand \enquote  [1]{``#1''}%
\providecommand \bibnamefont  [1]{#1}%
\providecommand \bibfnamefont [1]{#1}%
\providecommand \citenamefont [1]{#1}%
\providecommand \href@noop [0]{\@secondoftwo}%
\providecommand \href [0]{\begingroup \@sanitize@url \@href}%
\providecommand \@href[1]{\@@startlink{#1}\@@href}%
\providecommand \@@href[1]{\endgroup#1\@@endlink}%
\providecommand \@sanitize@url [0]{\catcode `\\12\catcode `\$12\catcode
  `\&12\catcode `\#12\catcode `\^12\catcode `\_12\catcode `\%12\relax}%
\providecommand \@@startlink[1]{}%
\providecommand \@@endlink[0]{}%
\providecommand \url  [0]{\begingroup\@sanitize@url \@url }%
\providecommand \@url [1]{\endgroup\@href {#1}{\urlprefix }}%
\providecommand \urlprefix  [0]{URL }%
\providecommand \Eprint [0]{\href }%
\providecommand \doibase [0]{http://dx.doi.org/}%
\providecommand \selectlanguage [0]{\@gobble}%
\providecommand \bibinfo  [0]{\@secondoftwo}%
\providecommand \bibfield  [0]{\@secondoftwo}%
\providecommand \translation [1]{[#1]}%
\providecommand \BibitemOpen [0]{}%
\providecommand \bibitemStop [0]{}%
\providecommand \bibitemNoStop [0]{.\EOS\space}%
\providecommand \EOS [0]{\spacefactor3000\relax}%
\providecommand \BibitemShut  [1]{\csname bibitem#1\endcsname}%
\let\auto@bib@innerbib\@empty
\bibitem [{\citenamefont {{Wolf, S. A. and Awschalom, D. D. and Buhrman, R. A.
  and Daughton, J. M. and von Moln{\'{a}}r, S. and Roukes, M. L. and
  Chtchelkanova, A. Y. and Treger}}(2001)}]{Wolf2001}%
  \BibitemOpen
  \bibfield  {author} {\bibinfo {author} {\bibfnamefont {D.~M.}\ \bibnamefont
  {{Wolf, S. A. and Awschalom, D. D. and Buhrman, R. A. and Daughton, J. M. and
  von Moln{\'{a}}r, S. and Roukes, M. L. and Chtchelkanova, A. Y. and
  Treger}}},\ }\href {\doibase 10.1126/science.1065389} {\bibfield  {journal}
  {\bibinfo  {journal} {Science}\ }\textbf {\bibinfo {volume} {294}},\ \bibinfo
  {pages} {1488} (\bibinfo {year} {2001})}\BibitemShut {NoStop}%
\bibitem [{\citenamefont {Dyakonov}\ and\ \citenamefont
  {Perel}(1971)}]{Dyakonov1971}%
  \BibitemOpen
  \bibfield  {author} {\bibinfo {author} {\bibfnamefont {M.~I.}\ \bibnamefont
  {Dyakonov}}\ and\ \bibinfo {author} {\bibfnamefont {V.~I.}\ \bibnamefont
  {Perel}},\ }\href {\doibase 10.1016/0375-9601(71)90196-4} {\bibfield
  {journal} {\bibinfo  {journal} {Phys. Lett. A}\ }\textbf {\bibinfo {volume}
  {35}},\ \bibinfo {pages} {459} (\bibinfo {year} {1971})}\BibitemShut
  {NoStop}%
\bibitem [{\citenamefont {Hirsch}(1999)}]{Hirsch1999}%
  \BibitemOpen
  \bibfield  {author} {\bibinfo {author} {\bibfnamefont {J.~E.}\ \bibnamefont
  {Hirsch}},\ }\href {\doibase 10.1103/PhysRevLett.83.1834} {\bibfield
  {journal} {\bibinfo  {journal} {Phys. Rev. Lett.}\ }\textbf {\bibinfo
  {volume} {83}},\ \bibinfo {pages} {1834} (\bibinfo {year} {1999})},\ \Eprint
  {http://arxiv.org/abs/0602330} {arXiv:0602330 [cond-mat]} \BibitemShut
  {NoStop}%
\bibitem [{\citenamefont {Zhang}(2000)}]{Zhang2000}%
  \BibitemOpen
  \bibfield  {author} {\bibinfo {author} {\bibfnamefont {S.}~\bibnamefont
  {Zhang}},\ }\href {\doibase 10.1103/PhysRevLett.85.393} {\bibfield  {journal}
  {\bibinfo  {journal} {Phys. Rev. Lett.}\ }\textbf {\bibinfo {volume} {85}},\
  \bibinfo {pages} {393} (\bibinfo {year} {2000})}\BibitemShut {NoStop}%
\bibitem [{\citenamefont {Murakami}\ \emph {et~al.}(2003)\citenamefont
  {Murakami}, \citenamefont {Nagaosa},\ and\ \citenamefont
  {Zhang}}]{Murakami2003}%
  \BibitemOpen
  \bibfield  {author} {\bibinfo {author} {\bibfnamefont {S.}~\bibnamefont
  {Murakami}}, \bibinfo {author} {\bibfnamefont {N.}~\bibnamefont {Nagaosa}}, \
  and\ \bibinfo {author} {\bibfnamefont {S.-c.}\ \bibnamefont {Zhang}},\ }\href
  {\doibase 10.1126/science.1087128} {\bibfield  {journal} {\bibinfo  {journal}
  {Science (80-. ).}\ }\textbf {\bibinfo {volume} {301}},\ \bibinfo {pages}
  {1348} (\bibinfo {year} {2003})}\BibitemShut {NoStop}%
\bibitem [{\citenamefont {Sinova}\ \emph {et~al.}(2004)\citenamefont {Sinova},
  \citenamefont {Culcer}, \citenamefont {Niu}, \citenamefont {Sinitsyn},
  \citenamefont {Jungwirth},\ and\ \citenamefont {MacDonald}}]{Sinova2004}%
  \BibitemOpen
  \bibfield  {author} {\bibinfo {author} {\bibfnamefont {J.}~\bibnamefont
  {Sinova}}, \bibinfo {author} {\bibfnamefont {D.}~\bibnamefont {Culcer}},
  \bibinfo {author} {\bibfnamefont {Q.}~\bibnamefont {Niu}}, \bibinfo {author}
  {\bibfnamefont {N.~A.}\ \bibnamefont {Sinitsyn}}, \bibinfo {author}
  {\bibfnamefont {T.}~\bibnamefont {Jungwirth}}, \ and\ \bibinfo {author}
  {\bibfnamefont {A.~H.}\ \bibnamefont {MacDonald}},\ }\href {\doibase
  10.1103/PhysRevLett.92.126603} {\bibfield  {journal} {\bibinfo  {journal}
  {Phys. Rev. Lett.}\ }\textbf {\bibinfo {volume} {92}},\ \bibinfo {pages}
  {126603} (\bibinfo {year} {2004})},\ \Eprint {http://arxiv.org/abs/0307663}
  {arXiv:0307663 [cond-mat]} \BibitemShut {NoStop}%
\bibitem [{\citenamefont {Schliemann}\ and\ \citenamefont
  {Loss}(2005)}]{Schliemann2005}%
  \BibitemOpen
  \bibfield  {author} {\bibinfo {author} {\bibfnamefont {J.}~\bibnamefont
  {Schliemann}}\ and\ \bibinfo {author} {\bibfnamefont {D.}~\bibnamefont
  {Loss}},\ }\href {\doibase 10.1103/PhysRevB.71.085308} {\bibfield  {journal}
  {\bibinfo  {journal} {Phys. Rev. B - Condens. Matter Mater. Phys.}\ }\textbf
  {\bibinfo {volume} {71}},\ \bibinfo {pages} {1} (\bibinfo {year} {2005})},\
  \Eprint {http://arxiv.org/abs/0405436} {arXiv:0405436 [cond-mat]}
  \BibitemShut {NoStop}%
\bibitem [{\citenamefont {Dresselhaus}(1955)}]{dresselhaus1955}%
  \BibitemOpen
  \bibfield  {author} {\bibinfo {author} {\bibfnamefont {G.}~\bibnamefont
  {Dresselhaus}},\ }\href {\doibase 10.1103/PhysRev.100.580} {\bibfield
  {journal} {\bibinfo  {journal} {Phys.Rev.}\ }\textbf {\bibinfo {volume}
  {100}},\ \bibinfo {pages} {580} (\bibinfo {year} {1955})},\ \Eprint
  {http://arxiv.org/abs/arXiv:1011.1669v3} {arXiv:arXiv:1011.1669v3}
  \BibitemShut {NoStop}%
\bibitem [{\citenamefont {Bychkov}\ and\ \citenamefont
  {Rashba}(1984)}]{Bychkov1984}%
  \BibitemOpen
  \bibfield  {author} {\bibinfo {author} {\bibfnamefont {Y.~a.}\ \bibnamefont
  {Bychkov}}\ and\ \bibinfo {author} {\bibfnamefont {E.~I.}\ \bibnamefont
  {Rashba}},\ }\href {\doibase 10.1088/0022-3719/17/33/015} {\bibfield
  {journal} {\bibinfo  {journal} {J. Phys. C Solid State Phys.}\ }\textbf
  {\bibinfo {volume} {17}},\ \bibinfo {pages} {6039} (\bibinfo {year}
  {1984})}\BibitemShut {NoStop}%
\bibitem [{\citenamefont {Reichl}(2014)}]{Reichl2014}%
  \BibitemOpen
  \bibfield  {author} {\bibinfo {author} {\bibfnamefont {C.}~\bibnamefont
  {Reichl}},\ }\emph {\bibinfo {title} {MBE growth of ultrahigh-mobility 2DEGs
  in GaAs/AlGaAs}},\ \href@noop {} {Ph.D. thesis},\ \bibinfo  {school} {ETH
  Z\"urich} (\bibinfo {year} {2014})\BibitemShut {NoStop}%
\bibitem [{\citenamefont {Manfra}(2014)}]{Manfra2014}%
  \BibitemOpen
  \bibfield  {author} {\bibinfo {author} {\bibfnamefont {M.}~\bibnamefont
  {Manfra}},\ }\href@noop {} {\bibfield  {journal} {\bibinfo  {journal} {Annu.
  Rev. Condens. Matter Phys.}\ } (\bibinfo {year} {2014})}\BibitemShut
  {NoStop}%
\bibitem [{\citenamefont {Hirmer}(2013)}]{Hirmer2013}%
  \BibitemOpen
  \bibfield  {author} {\bibinfo {author} {\bibfnamefont {M.}~\bibnamefont
  {Hirmer}},\ }\emph {\bibinfo {title} {High-mobility two-dimensional hole
  gases in III-V semiconductor heterostructures: growth and transport
  properties}},\ \href@noop {} {Ph.D. thesis},\ \bibinfo  {school}
  {Universit\"at Regensburg} (\bibinfo {year} {2013})\BibitemShut {NoStop}%
\bibitem [{\citenamefont {Watson}\ \emph {et~al.}(2012)\citenamefont {Watson},
  \citenamefont {Mondal}, \citenamefont {Gardner}, \citenamefont {Csáthy},\
  and\ \citenamefont {Manfra}}]{Watson2012}%
  \BibitemOpen
  \bibfield  {author} {\bibinfo {author} {\bibfnamefont {J.}~\bibnamefont
  {Watson}}, \bibinfo {author} {\bibfnamefont {S.}~\bibnamefont {Mondal}},
  \bibinfo {author} {\bibfnamefont {G.}~\bibnamefont {Gardner}}, \bibinfo
  {author} {\bibfnamefont {G.}~\bibnamefont {Csáthy}}, \ and\ \bibinfo
  {author} {\bibfnamefont {M.}~\bibnamefont {Manfra}},\ }\href@noop {}
  {\bibfield  {journal} {\bibinfo  {journal} {Phys. Rev. B}\ }\textbf {\bibinfo
  {volume} {85}},\ \bibinfo {pages} {165301} (\bibinfo {year}
  {2012})}\BibitemShut {NoStop}%
\bibitem [{\citenamefont {Kato}\ \emph {et~al.}(2004)\citenamefont {Kato},
  \citenamefont {Myers}, \citenamefont {Gossard},\ and\ \citenamefont
  {Awschalom}}]{Kato2004}%
  \BibitemOpen
  \bibfield  {author} {\bibinfo {author} {\bibfnamefont {Y.~K.}\ \bibnamefont
  {Kato}}, \bibinfo {author} {\bibfnamefont {R.~C.}\ \bibnamefont {Myers}},
  \bibinfo {author} {\bibfnamefont {A.~C.}\ \bibnamefont {Gossard}}, \ and\
  \bibinfo {author} {\bibfnamefont {D.~D.}\ \bibnamefont {Awschalom}},\ }\href
  {\doibase 10.1126/science.1105514} {\bibfield  {journal} {\bibinfo  {journal}
  {Science (80-. ).}\ }\textbf {\bibinfo {volume} {306}},\ \bibinfo {pages}
  {1910} (\bibinfo {year} {2004})}\BibitemShut {NoStop}%
\bibitem [{\citenamefont {Sih}\ \emph {et~al.}(2005)\citenamefont {Sih},
  \citenamefont {Myers}, \citenamefont {Kato}, \citenamefont {Lau},
  \citenamefont {Gossard},\ and\ \citenamefont {Awschalom}}]{Sih2005}%
  \BibitemOpen
  \bibfield  {author} {\bibinfo {author} {\bibfnamefont {V.}~\bibnamefont
  {Sih}}, \bibinfo {author} {\bibfnamefont {R.~C.}\ \bibnamefont {Myers}},
  \bibinfo {author} {\bibfnamefont {Y.~K.}\ \bibnamefont {Kato}}, \bibinfo
  {author} {\bibfnamefont {W.~H.}\ \bibnamefont {Lau}}, \bibinfo {author}
  {\bibfnamefont {A.~C.}\ \bibnamefont {Gossard}}, \ and\ \bibinfo {author}
  {\bibfnamefont {D.~D.}\ \bibnamefont {Awschalom}},\ }\href {\doibase
  10.1038/nphys009} {\bibfield  {journal} {\bibinfo  {journal} {Nat. Phys.}\
  }\textbf {\bibinfo {volume} {1}},\ \bibinfo {pages} {31} (\bibinfo {year}
  {2005})},\ \Eprint {http://arxiv.org/abs/0506704} {arXiv:0506704 [cond-mat]}
  \BibitemShut {NoStop}%
\bibitem [{\citenamefont {Wunderlich}\ \emph {et~al.}(2005)\citenamefont
  {Wunderlich}, \citenamefont {Kaestner}, \citenamefont {Sinova},\ and\
  \citenamefont {Jungwirth}}]{Wunderlich2005}%
  \BibitemOpen
  \bibfield  {author} {\bibinfo {author} {\bibfnamefont {J.}~\bibnamefont
  {Wunderlich}}, \bibinfo {author} {\bibfnamefont {B.}~\bibnamefont
  {Kaestner}}, \bibinfo {author} {\bibfnamefont {J.}~\bibnamefont {Sinova}}, \
  and\ \bibinfo {author} {\bibfnamefont {T.}~\bibnamefont {Jungwirth}},\ }\href
  {\doibase 10.1103/PhysRevLett.94.047204} {\bibfield  {journal} {\bibinfo
  {journal} {Phys. Rev. Lett.}\ }\textbf {\bibinfo {volume} {94}},\ \bibinfo
  {pages} {1} (\bibinfo {year} {2005})},\ \Eprint
  {http://arxiv.org/abs/0410295} {arXiv:0410295 [cond-mat]} \BibitemShut
  {NoStop}%
\bibitem [{\citenamefont {Hernandez}\ \emph {et~al.}(2013)\citenamefont
  {Hernandez}, \citenamefont {Nunes}, \citenamefont {Gusev},\ and\
  \citenamefont {Bakarov}}]{Hernandez2013}%
  \BibitemOpen
  \bibfield  {author} {\bibinfo {author} {\bibfnamefont {F.~G.~G.}\
  \bibnamefont {Hernandez}}, \bibinfo {author} {\bibfnamefont {L.~M.}\
  \bibnamefont {Nunes}}, \bibinfo {author} {\bibfnamefont {G.~M.}\ \bibnamefont
  {Gusev}}, \ and\ \bibinfo {author} {\bibfnamefont {A.~K.}\ \bibnamefont
  {Bakarov}},\ }\href {\doibase 10.1103/PhysRevB.88.161305} {\bibfield
  {journal} {\bibinfo  {journal} {Phys. Rev. B}\ }\textbf {\bibinfo {volume}
  {88}},\ \bibinfo {pages} {161305} (\bibinfo {year} {2013})}\BibitemShut
  {NoStop}%
\bibitem [{\citenamefont {Miller}(1985)}]{Miller1985}%
  \BibitemOpen
  \bibfield  {author} {\bibinfo {author} {\bibfnamefont {D.~L.}\ \bibnamefont
  {Miller}},\ }\href {\doibase 10.1063/1.96262} {\bibfield  {journal} {\bibinfo
   {journal} {Appl. Phys. Lett.}\ }\textbf {\bibinfo {volume} {47}},\ \bibinfo
  {pages} {1309} (\bibinfo {year} {1985})}\BibitemShut {NoStop}%
\bibitem [{\citenamefont {Pfeiffer}\ \emph {et~al.}(1990)\citenamefont
  {Pfeiffer}, \citenamefont {West}, \citenamefont {Stormer}, \citenamefont
  {Eisenstein}, \citenamefont {Baldwin}, \citenamefont {Gershoni},\ and\
  \citenamefont {Spector}}]{Pfeiffer1990}%
  \BibitemOpen
  \bibfield  {author} {\bibinfo {author} {\bibfnamefont {L.}~\bibnamefont
  {Pfeiffer}}, \bibinfo {author} {\bibfnamefont {K.~W.}\ \bibnamefont {West}},
  \bibinfo {author} {\bibfnamefont {H.~L.}\ \bibnamefont {Stormer}}, \bibinfo
  {author} {\bibfnamefont {J.~P.}\ \bibnamefont {Eisenstein}}, \bibinfo
  {author} {\bibfnamefont {K.~W.}\ \bibnamefont {Baldwin}}, \bibinfo {author}
  {\bibfnamefont {D.}~\bibnamefont {Gershoni}}, \ and\ \bibinfo {author}
  {\bibfnamefont {J.}~\bibnamefont {Spector}},\ }\href {\doibase
  10.1063/1.103121} {\bibfield  {journal} {\bibinfo  {journal} {Appl. Phys.
  Lett.}\ }\textbf {\bibinfo {volume} {56}},\ \bibinfo {pages} {1697} (\bibinfo
  {year} {1990})}\BibitemShut {NoStop}%
\bibitem [{\citenamefont {Pfeiffer}\ \emph {et~al.}(1993)\citenamefont
  {Pfeiffer}, \citenamefont {St{\"{o}}rmer}, \citenamefont {Baldwin},
  \citenamefont {West}, \citenamefont {Go{\~{n}}i}, \citenamefont {Pinczuk},
  \citenamefont {Ashoori}, \citenamefont {Dignam},\ and\ \citenamefont
  {Wegscheider}}]{Pfeiffer1993}%
  \BibitemOpen
  \bibfield  {author} {\bibinfo {author} {\bibfnamefont {L.}~\bibnamefont
  {Pfeiffer}}, \bibinfo {author} {\bibfnamefont {H.}~\bibnamefont
  {St{\"{o}}rmer}}, \bibinfo {author} {\bibfnamefont {K.}~\bibnamefont
  {Baldwin}}, \bibinfo {author} {\bibfnamefont {K.}~\bibnamefont {West}},
  \bibinfo {author} {\bibfnamefont {A.}~\bibnamefont {Go{\~{n}}i}}, \bibinfo
  {author} {\bibfnamefont {A.}~\bibnamefont {Pinczuk}}, \bibinfo {author}
  {\bibfnamefont {R.}~\bibnamefont {Ashoori}}, \bibinfo {author} {\bibfnamefont
  {M.}~\bibnamefont {Dignam}}, \ and\ \bibinfo {author} {\bibfnamefont
  {W.}~\bibnamefont {Wegscheider}},\ }\href {\doibase
  10.1016/0022-0248(93)90746-J} {\bibfield  {journal} {\bibinfo  {journal} {J.
  Cryst. Growth}\ }\textbf {\bibinfo {volume} {127}},\ \bibinfo {pages} {849}
  (\bibinfo {year} {1993})}\BibitemShut {NoStop}%
\bibitem [{\citenamefont {Giglberger}\ \emph {et~al.}(2007)\citenamefont
  {Giglberger}, \citenamefont {Golub}, \citenamefont {Bel'kov}, \citenamefont
  {Danilov}, \citenamefont {Schuh}, \citenamefont {Gerl}, \citenamefont
  {Rohlfing}, \citenamefont {Stahl}, \citenamefont {Wegscheider}, \citenamefont
  {Weiss}, \citenamefont {Prettl},\ and\ \citenamefont
  {Ganichev}}]{giglberger2007}%
  \BibitemOpen
  \bibfield  {author} {\bibinfo {author} {\bibfnamefont {S.}~\bibnamefont
  {Giglberger}}, \bibinfo {author} {\bibfnamefont {L.~E.}\ \bibnamefont
  {Golub}}, \bibinfo {author} {\bibfnamefont {V.~V.}\ \bibnamefont {Bel'kov}},
  \bibinfo {author} {\bibfnamefont {S.~N.}\ \bibnamefont {Danilov}}, \bibinfo
  {author} {\bibfnamefont {D.}~\bibnamefont {Schuh}}, \bibinfo {author}
  {\bibfnamefont {C.}~\bibnamefont {Gerl}}, \bibinfo {author} {\bibfnamefont
  {F.}~\bibnamefont {Rohlfing}}, \bibinfo {author} {\bibfnamefont
  {J.}~\bibnamefont {Stahl}}, \bibinfo {author} {\bibfnamefont
  {W.}~\bibnamefont {Wegscheider}}, \bibinfo {author} {\bibfnamefont
  {D.}~\bibnamefont {Weiss}}, \bibinfo {author} {\bibfnamefont
  {W.}~\bibnamefont {Prettl}}, \ and\ \bibinfo {author} {\bibfnamefont {S.~D.}\
  \bibnamefont {Ganichev}},\ }\href {\doibase 10.1103/PhysRevB.75.035327}
  {\bibfield  {journal} {\bibinfo  {journal} {Phys. Rev. B - Condens. Matter
  Mater. Phys.}\ }\textbf {\bibinfo {volume} {75}},\ \bibinfo {pages} {1}
  (\bibinfo {year} {2007})},\ \Eprint {http://arxiv.org/abs/0609569}
  {arXiv:0609569 [cond-mat]} \BibitemShut {NoStop}%
\bibitem [{\citenamefont {Riedi}\ \emph {et~al.}(2016)\citenamefont {Riedi},
  \citenamefont {Reichl}, \citenamefont {Berl}, \citenamefont {Alt},
  \citenamefont {Maier},\ and\ \citenamefont {Wegscheider}}]{Riedi2016}%
  \BibitemOpen
  \bibfield  {author} {\bibinfo {author} {\bibfnamefont {S.}~\bibnamefont
  {Riedi}}, \bibinfo {author} {\bibfnamefont {C.}~\bibnamefont {Reichl}},
  \bibinfo {author} {\bibfnamefont {M.}~\bibnamefont {Berl}}, \bibinfo {author}
  {\bibfnamefont {L.}~\bibnamefont {Alt}}, \bibinfo {author} {\bibfnamefont
  {A.}~\bibnamefont {Maier}}, \ and\ \bibinfo {author} {\bibfnamefont
  {W.}~\bibnamefont {Wegscheider}},\ }\href {\doibase
  10.1016/j.jcrysgro.2016.09.022} {\bibfield  {journal} {\bibinfo  {journal}
  {J. Cryst. Growth}\ }\textbf {\bibinfo {volume} {455}},\ \bibinfo {pages}
  {37} (\bibinfo {year} {2016})}\BibitemShut {NoStop}%
\bibitem [{\citenamefont {Harrison}\ and\ \citenamefont
  {Valavanis}(2016)}]{Harrison2016}%
  \BibitemOpen
  \bibfield  {author} {\bibinfo {author} {\bibfnamefont {P.}~\bibnamefont
  {Harrison}}\ and\ \bibinfo {author} {\bibfnamefont {A.}~\bibnamefont
  {Valavanis}},\ }\href {\doibase 10.1002/9781118923337} {\emph {\bibinfo
  {title} {Quantum Wells, Wires Dots}}}\ (\bibinfo  {publisher} {John Wiley
  {\&} Sons, Ltd},\ \bibinfo {address} {Chichester, UK},\ \bibinfo {year}
  {2016})\BibitemShut {NoStop}%
\bibitem [{\citenamefont {Birner}()}]{Nextnano}%
  \BibitemOpen
  \bibfield  {author} {\bibinfo {author} {\bibfnamefont {S.}~\bibnamefont
  {Birner}},\ }\href {www.nextnano.de} {\enquote {\bibinfo {title}
  {{Nextnano}},}\ }\bibinfo {howpublished} {www.nextnano.de}\BibitemShut
  {NoStop}%
\bibitem [{\citenamefont {Winkler}\ \emph {et~al.}(2008)\citenamefont
  {Winkler}, \citenamefont {Culcer}, \citenamefont {Papadakis}, \citenamefont
  {Habib},\ and\ \citenamefont {Shayegan}}]{winkler2008}%
  \BibitemOpen
  \bibfield  {author} {\bibinfo {author} {\bibfnamefont {R.}~\bibnamefont
  {Winkler}}, \bibinfo {author} {\bibfnamefont {D.}~\bibnamefont {Culcer}},
  \bibinfo {author} {\bibfnamefont {S.~J.}\ \bibnamefont {Papadakis}}, \bibinfo
  {author} {\bibfnamefont {B.}~\bibnamefont {Habib}}, \ and\ \bibinfo {author}
  {\bibfnamefont {M.}~\bibnamefont {Shayegan}},\ }\href {\doibase
  10.1088/0268-1242/23/11/114017} {\bibfield  {journal} {\bibinfo  {journal}
  {Semicond. Sci. Technol.}\ }\textbf {\bibinfo {volume} {23}},\ \bibinfo
  {pages} {114017} (\bibinfo {year} {2008})},\ \Eprint
  {http://arxiv.org/abs/0811.1744} {arXiv:0811.1744} \BibitemShut {NoStop}%
\bibitem [{\citenamefont {Nomura}\ \emph
  {et~al.}(2005{\natexlab{a}})\citenamefont {Nomura}, \citenamefont
  {Wunderlich}, \citenamefont {Sinova}, \citenamefont {Kaestner}, \citenamefont
  {MacDonald},\ and\ \citenamefont {Jungwirth}}]{nomura2005b}%
  \BibitemOpen
  \bibfield  {author} {\bibinfo {author} {\bibfnamefont {K.}~\bibnamefont
  {Nomura}}, \bibinfo {author} {\bibfnamefont {J.}~\bibnamefont {Wunderlich}},
  \bibinfo {author} {\bibfnamefont {J.}~\bibnamefont {Sinova}}, \bibinfo
  {author} {\bibfnamefont {B.}~\bibnamefont {Kaestner}}, \bibinfo {author}
  {\bibfnamefont {A.~H.}\ \bibnamefont {MacDonald}}, \ and\ \bibinfo {author}
  {\bibfnamefont {T.}~\bibnamefont {Jungwirth}},\ }\href {\doibase
  10.1103/PhysRevB.72.245330} {\bibfield  {journal} {\bibinfo  {journal} {Phys.
  Rev. B - Condens. Matter Mater. Phys.}\ }\textbf {\bibinfo {volume} {72}},\
  \bibinfo {pages} {1} (\bibinfo {year} {2005}{\natexlab{a}})},\ \Eprint
  {http://arxiv.org/abs/0508532} {arXiv:0508532 [cond-mat]} \BibitemShut
  {NoStop}%
\bibitem [{\citenamefont {Nichele}(2014)}]{Nichele2014}%
  \BibitemOpen
  \bibfield  {author} {\bibinfo {author} {\bibfnamefont {F.}~\bibnamefont
  {Nichele}},\ }\emph {\bibinfo {title} {Transport experiments in
  two-dimensional systems with strong spin-orbit interaction}},\ \href@noop {}
  {Ph.D. thesis},\ \bibinfo  {school} {ETH Z\"urich} (\bibinfo {year}
  {2014})\BibitemShut {NoStop}%
\bibitem [{\citenamefont {Hankiewicz}\ \emph {et~al.}(2004)\citenamefont
  {Hankiewicz}, \citenamefont {Molenkamp}, \citenamefont {Jungwirth},\ and\
  \citenamefont {Sinova}}]{Hankiewicz2004}%
  \BibitemOpen
  \bibfield  {author} {\bibinfo {author} {\bibfnamefont {E.~M.}\ \bibnamefont
  {Hankiewicz}}, \bibinfo {author} {\bibfnamefont {L.~W.}\ \bibnamefont
  {Molenkamp}}, \bibinfo {author} {\bibfnamefont {T.}~\bibnamefont
  {Jungwirth}}, \ and\ \bibinfo {author} {\bibfnamefont {J.}~\bibnamefont
  {Sinova}},\ }\href {\doibase 10.1103/PhysRevB.70.241301} {\bibfield
  {journal} {\bibinfo  {journal} {Phys. Rev. B - Condens. Matter Mater. Phys.}\
  }\textbf {\bibinfo {volume} {70}},\ \bibinfo {pages} {1} (\bibinfo {year}
  {2004})},\ \Eprint {http://arxiv.org/abs/0409334} {arXiv:0409334 [cond-mat]}
  \BibitemShut {NoStop}%
\bibitem [{\citenamefont {Nikoli{\'{c}}}\ \emph {et~al.}(2005)\citenamefont
  {Nikoli{\'{c}}}, \citenamefont {Z{\^{a}}rbo},\ and\ \citenamefont
  {Souma}}]{nikolic2005}%
  \BibitemOpen
  \bibfield  {author} {\bibinfo {author} {\bibfnamefont {B.~K.}\ \bibnamefont
  {Nikoli{\'{c}}}}, \bibinfo {author} {\bibfnamefont {L.~P.}\ \bibnamefont
  {Z{\^{a}}rbo}}, \ and\ \bibinfo {author} {\bibfnamefont {S.}~\bibnamefont
  {Souma}},\ }\href {\doibase 10.1103/PhysRevB.72.075361} {\bibfield  {journal}
  {\bibinfo  {journal} {Phys. Rev. B}\ }\textbf {\bibinfo {volume} {72}},\
  \bibinfo {pages} {075361} (\bibinfo {year} {2005})},\ \Eprint
  {http://arxiv.org/abs/0408693} {arXiv:0408693 [arXiv:cond-mat]} \BibitemShut
  {NoStop}%
\bibitem [{\citenamefont {Nomura}\ \emph
  {et~al.}(2005{\natexlab{b}})\citenamefont {Nomura}, \citenamefont {Sinova},
  \citenamefont {Jungwirth}, \citenamefont {Niu},\ and\ \citenamefont
  {MacDonald}}]{nomura2005d}%
  \BibitemOpen
  \bibfield  {author} {\bibinfo {author} {\bibfnamefont {K.}~\bibnamefont
  {Nomura}}, \bibinfo {author} {\bibfnamefont {J.}~\bibnamefont {Sinova}},
  \bibinfo {author} {\bibfnamefont {T.}~\bibnamefont {Jungwirth}}, \bibinfo
  {author} {\bibfnamefont {Q.}~\bibnamefont {Niu}}, \ and\ \bibinfo {author}
  {\bibfnamefont {A.~H.}\ \bibnamefont {MacDonald}},\ }\href {\doibase
  10.1103/PhysRevB.71.041304} {\bibfield  {journal} {\bibinfo  {journal} {Phys.
  Rev. B - Condens. Matter Mater. Phys.}\ }\textbf {\bibinfo {volume} {71}},\
  \bibinfo {pages} {1} (\bibinfo {year} {2005}{\natexlab{b}})},\ \Eprint
  {http://arxiv.org/abs/0407279} {arXiv:0407279 [cond-mat]} \BibitemShut
  {NoStop}%
\bibitem [{\citenamefont {Schl\"apfer}\ \emph {et~al.}(2016)\citenamefont
  {Schl\"apfer}, \citenamefont {Dietsche}, \citenamefont {Reichl},\ and\
  \citenamefont {Wegscheider}}]{Schlaepfer2016}%
  \BibitemOpen
  \bibfield  {author} {\bibinfo {author} {\bibfnamefont {F.}~\bibnamefont
  {Schl\"apfer}}, \bibinfo {author} {\bibfnamefont {W.}~\bibnamefont
  {Dietsche}}, \bibinfo {author} {\bibfnamefont {C.}~\bibnamefont {Reichl}}, \
  and\ \bibinfo {author} {\bibfnamefont {W.}~\bibnamefont {Wegscheider}},\
  }\href@noop {} {\bibfield  {journal} {\bibinfo  {journal} {J. Cryst. Growth}\
  } (\bibinfo {year} {2016})}\BibitemShut {NoStop}%
\bibitem [{\citenamefont {Gammon}\ \emph {et~al.}(1996)\citenamefont {Gammon},
  \citenamefont {Snow}, \citenamefont {Shanabrook}, \citenamefont {Katzer},\
  and\ \citenamefont {Park}}]{gammon1996}%
  \BibitemOpen
  \bibfield  {author} {\bibinfo {author} {\bibfnamefont {D.}~\bibnamefont
  {Gammon}}, \bibinfo {author} {\bibfnamefont {E.~S.}\ \bibnamefont {Snow}},
  \bibinfo {author} {\bibfnamefont {B.~V.}\ \bibnamefont {Shanabrook}},
  \bibinfo {author} {\bibfnamefont {D.~S.}\ \bibnamefont {Katzer}}, \ and\
  \bibinfo {author} {\bibfnamefont {D.}~\bibnamefont {Park}},\ }\href {\doibase
  10.1103/PhysRevLett.76.3005} {\bibfield  {journal} {\bibinfo  {journal}
  {Phys. Rev. Lett.}\ }\textbf {\bibinfo {volume} {76}},\ \bibinfo {pages}
  {3005} (\bibinfo {year} {1996})}\BibitemShut {NoStop}%
\bibitem [{\citenamefont {{Van Kesteren}}\ \emph {et~al.}(1990)\citenamefont
  {{Van Kesteren}}, \citenamefont {Cosman}, \citenamefont {{Van Der Poel}},\
  and\ \citenamefont {Foxon}}]{vankesteren1990a}%
  \BibitemOpen
  \bibfield  {author} {\bibinfo {author} {\bibfnamefont {H.~W.}\ \bibnamefont
  {{Van Kesteren}}}, \bibinfo {author} {\bibfnamefont {E.~C.}\ \bibnamefont
  {Cosman}}, \bibinfo {author} {\bibfnamefont {W.~A. J.~A.}\ \bibnamefont {{Van
  Der Poel}}}, \ and\ \bibinfo {author} {\bibfnamefont {C.~T.}\ \bibnamefont
  {Foxon}},\ }\href {\doibase 10.1103/PhysRevB.41.5283} {\bibfield  {journal}
  {\bibinfo  {journal} {Phys. Rev. B}\ }\textbf {\bibinfo {volume} {41}},\
  \bibinfo {pages} {5283} (\bibinfo {year} {1990})}\BibitemShut {NoStop}%
\bibitem [{\citenamefont {Blackwood}\ \emph {et~al.}(1994)\citenamefont
  {Blackwood}, \citenamefont {Snelling}, \citenamefont {Harley}, \citenamefont
  {Andrews},\ and\ \citenamefont {Foxon}}]{blackwood1994}%
  \BibitemOpen
  \bibfield  {author} {\bibinfo {author} {\bibfnamefont {E.}~\bibnamefont
  {Blackwood}}, \bibinfo {author} {\bibfnamefont {M.~J.}\ \bibnamefont
  {Snelling}}, \bibinfo {author} {\bibfnamefont {R.~T.}\ \bibnamefont
  {Harley}}, \bibinfo {author} {\bibfnamefont {S.~R.}\ \bibnamefont {Andrews}},
  \ and\ \bibinfo {author} {\bibfnamefont {C.~T.~B.}\ \bibnamefont {Foxon}},\
  }\href {\doibase 10.1103/PhysRevB.50.14246} {\bibfield  {journal} {\bibinfo
  {journal} {Phys. Rev. B}\ }\textbf {\bibinfo {volume} {50}},\ \bibinfo
  {pages} {14246} (\bibinfo {year} {1994})}\BibitemShut {NoStop}%
\bibitem [{\citenamefont {van Exter}\ \emph {et~al.}(1998)\citenamefont {van
  Exter}, \citenamefont {Willemsen},\ and\ \citenamefont
  {Woerdman}}]{vanexter1998}%
  \BibitemOpen
  \bibfield  {author} {\bibinfo {author} {\bibfnamefont {M.}~\bibnamefont {van
  Exter}}, \bibinfo {author} {\bibfnamefont {M.}~\bibnamefont {Willemsen}}, \
  and\ \bibinfo {author} {\bibfnamefont {J.}~\bibnamefont {Woerdman}},\ }\href
  {\doibase 10.1103/PhysRevA.58.4191} {\bibfield  {journal} {\bibinfo
  {journal} {Phys. Rev. A}\ }\textbf {\bibinfo {volume} {58}},\ \bibinfo
  {pages} {4191} (\bibinfo {year} {1998})}\BibitemShut {NoStop}%
\end{thebibliography}%

\end{document}